\documentclass[prd,superscriptaddress,showpacs,nofootinbib,amsmath,amssymb,aps,10pt]{revtex4}

\usepackage{bm}
\usepackage{amsfonts}
\usepackage{latexsym}
\usepackage[latin1]{inputenc}
\usepackage{graphicx}
\usepackage{amsmath}
\usepackage{palatino}
\usepackage{mathpazo}
\usepackage{booktabs}
\usepackage{dcolumn}

\usepackage{wrapfig}
\usepackage{wasysym}
\usepackage[outdir=./]{epstopdf}

\def\jnl@style{\it}
\def\aaref@jnl#1{{\jnl@style#1}}

\def\aaref@jnl#1{{\jnl@style#1}}

\def\aj{\aaref@jnl{AJ}}                   
\def\apj{\aaref@jnl{ApJ}}                 
\def\apjl{\aaref@jnl{ApJ}}                
\def\apjs{\aaref@jnl{ApJS}}               
\def\apss{\aaref@jnl{Ap\&SS}}             
\def\aap{\aaref@jnl{A\&A}}                
\def\aapr{\aaref@jnl{A\&A~Rev.}}          
\def\aaps{\aaref@jnl{A\&AS}}              
\def\mnras{\aaref@jnl{Mon.~Not.~Roy.~Astron.~Soc.}}             
\def\prd{\aaref@jnl{Phys.~Rev.~D}}        
\def\prc{\aaref@jnl{Phys.~Rev.~C}}  
\def\prl{\aaref@jnl{Phys.~Rev.~Lett.}}    
\def\qjras{\aaref@jnl{QJRAS}}             
\def\skytel{\aaref@jnl{S\&T}}             
\def\ssr{\aaref@jnl{Space~Sci.~Rev.}}     
\def\zap{\aaref@jnl{ZAp}}                 
\def\nat{\aaref@jnl{Nature}}              
\def\aplett{\aaref@jnl{Astrophys.~Lett.}} 
\def\apspr{\aaref@jnl{Astrophys.~Space~Phys.~Res.}} 
\def\physrep{\aaref@jnl{Phys.~Rep.}}      
\def\physscr{\aaref@jnl{Phys.~Scr}}       
\def\commat{\aaref@jnl{Comm.~Math.~Phys.}}              
\def\science{\aaref@jnl{Science}}               
\def\cqg{\aaref@jnl{Classical Quant.~Grav.}}            
\def\jpcs{\aaref@jnl{JPCS}}                                     
\def\ijmpd{\aaref@jnl{Int.~J.~Mod.~Phys.~D}}                    
\def\grg{\aaref@jnl{Gen.~Relat.~Gravit.}}               
\def\rpp{\aaref@jnl{Rep.~Prog.~Phys.}}          
\def\npa{\aaref@jnl{Nucl.~Phys.~A}}        
\def\lrr{\aaref@jnl{Living Rev.~Rel.}}                   
\def\jcap{\aaref@jnl{J.~Cosmology Astropart.~Phys.}}    
\def\rmp{\aaref@jnl{Rev.~Mod.~Phys.}}   

\allowdisplaybreaks[1]

\begin{document}

\title{Gravitational wave asteroseismology of neutron and strange stars in $R^2$  gravity}

\author{Kalin V. Staykov}
\email{kstaykov@phys.uni-sofia.bg}
\affiliation{Department of Theoretical Physics, Faculty of Physics, Sofia University, Sofia 1164, Bulgaria}

\author{Daniela D. Doneva}
\email{daniela.doneva@uni-tuebingen.de}
\affiliation{Theoretical Astrophysics, Eberhard Karls University of T\"ubingen, T\"ubingen 72076, Germany}
\affiliation{INRNE - Bulgarian Academy of Sciences, 1784  Sofia, Bulgaria}

\author{Stoytcho S. Yazadjiev}
\email{yazad@phys.uni-sofia.bg}
\affiliation{Department of Theoretical Physics, Faculty of Physics, Sofia University, Sofia 1164, Bulgaria}
\affiliation{Theoretical Astrophysics, Eberhard Karls University of T\"ubingen, T\"ubingen 72076, Germany}

\author{Kostas D. Kokkotas}
\email{kostas.kokkotas@uni-tuebingen.de}
\affiliation{Theoretical Astrophysics, Eberhard Karls University of T\"ubingen, T\"ubingen 72076, Germany}
\affiliation{Department of Physics, Aristotle University of Thessaloniki, Thessaloniki 54124, Greece}


\begin{abstract}

We study the oscillations of neutron and strange stars in $R^2$ gravity. More precisely the nonradial 
$f$-modes are examined and the differences with pure general relativity are investigated. 
Using these results we build several gravitational wave asteroseismology relations. Our goal is to determine up to what extend 
these relations are equation of state independent and whether they deviate enough from general relativity in order to produce an observable effect. 
The results show that the differences coming from $R^2$ gravity are up to 10\% and that will be difficult to be observed in the near future. 
On the other hand the small deviations in some of the asteroseismology relations show that they are not only equation of state independent, 
but they are also quite insensitive to the gravitational theory. That is why solving the inverse problem can give us quite robust estimates of the neutron star parameters.

\end{abstract}
\pacs{}
\maketitle
\date{}
\section{Introduction}

In the last few years the accelerated expansion of the universe has been confirmed, leading to the investigations of different cosmological models capable of explaining it. A new type of matter should be introduced when considering this problem in the context of general relativity (GR) -- the so-called dark energy which constitutes about 73\% of the total energy content of the Universe and exhibits some exotic properties, like a negative pressure-to-density ratio. A way of overcoming this is to explore the generalized theories of gravity as a possible alternative explanation for the accelerated expansion. One of the most popular representatives are the f(R) theories which allow us to  exclude the dark energy hypothesis. These theories of gravity are a natural generalization of GR, derived by replacing the Lagrangian in the Einstein-Hilbert action with a more general one \cite{Sotiriou_10,De Felice_10, Nojiri_11}.

The viability of any theory of gravity should be tested against the astrophysical observations too. In this paper we are investigating one such astrophysical application for a specific type of $f(R)$ theory, namely the so-called $R-$squared  gravity, where the form of the Lagrangian we adopt is $f(R) = R + aR^2$.
Differences between general relativity and alternative theories are expected to occur for strong gravitational fields, such as the ones created by different compact objects like neutron stars (NS), strange stars (SS) and black holes (BH). The scope of our studies is limited to the first two of them.

Taking in consideration that the new generation of gravitational wave detectors will start operating in the next few years, we extend the studies of compact stars in $f(R)$  gravity to gravitational wave asteroseismology. The neutron star perturbations can lead to nonradial oscillations,  which are a source of gravitational waves. It is expected that in the near feature the emitted gravitational waves will be  observed by the earth based detectors and the obtained information about the oscillation frequencies will reveal a new spectrum of research possibilities for astrophysics. The full development of gravitational wave asteroseimology will let us use optimally the gained information. The relations and the methodology that would allow us to obtain the neutron star parameters via the gravitational wave observations have been developed in the last two decades in many papers for GR \cite{A_K_06,A_K_08,Benhar_04,Lau_10,Chirenti_15} and for some alternative models \cite{Sotani_03,Sotani_09,Yazad_12,Doneva_12,Sotani_14,Krueger_14,Sotani_04,Sotani_2005,Sotani_10,Silva_2014}, and in the last few years even for rapidly rotating neutron stars \cite{Gaertig_11,Doneva_13}. 

More detailed comments on some of the above-mentioned papers, which our work is based on, should be made.
In \cite{A_K_08,Kokkotas_2001} the authors investigated the fundamental and the pressure mode oscillation frequencies and damping times as functions of the average density $\sqrt{M/R^3}$ of the stars or as functions of the compactness $M/R$ for equations of state (EOS) with different stiffness. They also introduced some empirical relations between these quantities that can be used to infer the neutron star mass and the radius, i.e. solve the inverse problem. Later those relations were reexamined for a set of newer EOS by Benhar et al. \cite{Benhar_04}. Lau et al. \cite{Lau_10} suggested a different normalization, namely to use the so-called ``effective compactness''  $\eta \equiv \sqrt{M^3/I}$, with $I$ being the moment of inertia, instead of the compactness $M/R$, motivated by the studies in \cite{Lattimer_05}. Using that parameter they introduced an empirical relation valid for both NS and SS, which turned out to be quite EOS independent. In a recent paper Chirenti et al. \cite{Chirenti_15} investigated oscillations of neutron stars in GR with all of the above mentioned relations. They commented on the EOS independence of these empirical relations. Other EOS independent relations connecting the neutron star moment of inertia, quadrupole moment and the tidal Love numbers were considered for example in \cite{Yagi_13}--\cite{Urbanec_13}.

For the  first time the oscillation frequencies of neutron stars in alternative theories of gravity, and more specifically in the case of a scalar-tensor theory (STT), were investigated  in \cite{Sotani_04,Sotani_2005}. There one can also find a detailed derivation of the perturbation equations for STT in the so-called Cowling approximation (the background metric and the scalar field are fixed).

The relations presented in the above-mentioned papers are EOS independent to a large extent. Such relations are a useful tool for overcoming the uncertainty naturally caused by the EOS and can be used to explore generalized theories of gravity.  This will be our main goal in the present paper, namely to explore the various asteroseismology relations in $R^2$ gravity and to quantify the deviations from general relativity. Such results can be potentially used when solving the inverse problem, i.e. obtaining the stellar parameters from the observed oscillation frequencies, in order to impose constraints on the gravitational theory. In the current paper we will concentrate on the nonradial oscillations for which the pressure is the restoring force. The scope of our study is limited to the fundamental modes (f-modes) that are the major object of investigation in the literature cited above. The reason is that the pressure modes (p-modes) are a much less efficient source of gravitational waves, so it is expected that the main contribution to observational data will be from f-mode oscillations.

It is important to point out that in some of the above-mentioned papers, and more precisely in the more complicated cases concerning rapid rotation and alternative theories of gravity, the Cowling approximation is used \cite{Sotani_04,Gaertig_11,Doneva_13}. It is known that there are differences between the results obtained  via the Cowling approximation and the full general relativistic results, and the deviation decreases with the increase of the compactness. However, this approximation is accurate enough for qualitative investigations  \cite{Gaertig_11,Doneva_13}. That is why we will employ it in our studies.

The structure of our paper is as follows: In Section II the reduced field equations describing neutron stars in $R^2$ gravity are presented, in Section III the equations for the perturbations are examined, in Section IV the results are presented and discussed. The paper ends with conclusions.

\section{Basic equations}

In this section we briefly  present the basic equations describing equilibrium neutron and strange star solutions  in $R^2$ gravity.  More details on this problem can be found in \cite{Yazad_14,Staykov_14}.

The $f(R)$ theories are described by the following action:

 \begin{eqnarray}
 S = \frac{1}{16 \pi G}\int d^4x\sqrt{-g}f(R) + S_{matter}(g_{\mu\nu},\chi),
 \end{eqnarray}
where $R$ is the scalar curvature with respect to the spacetime metric $g_{\mu\nu}$. The action for the matter fields, denoted by $\chi$, is separated in the term $S_{matter}$. The following inequalities have to be satisfied for the $f(R)$ theory  to be free of ghosts and tachyonic instabilities:

 \begin{eqnarray}
 \frac{d^2f}{dR^2}\geq 0,\;\;\; \frac{df}{dR}>0.
 \end{eqnarray}
 In the case of $R^2$ gravity, which we are considering, the above inequalities give $a\geq 0$ and $1+2aR \geq 0$.

A useful fact, that we will employ in our studies, is the mathematical equivalence between $f(R)$ theories and Brans-Dicke theory (with $\omega_{BD}=0$ and nonzero scalar field potential) given by the action

 \begin{eqnarray}
 S = \frac{1}{16 \pi G}\int d^4x\sqrt{-g}\left[\Phi R - U(\Phi)\right] + S_{matter}(g_{\mu\nu},\chi),
 \end{eqnarray}
 where $\Phi$ and the potential $U(\Phi)$ are defined as follow: $\Phi = \frac{df(R)}{dR}$ and $U(\Phi) = R\frac{df}{dR}-f(R)$. In the case of $R^2$ gravity $\Phi = 1 + 2aR$ and the Brans-Dicke potential is $U(\Phi) = \frac{1}{4a}\left(\Phi-1\right)^2$.
 
While the above action is in the so-called Jordan, or physical, frame, for mathematical simplicity it is useful to study the scalar-tensor theories 
in the so-called Einstein frame. The metric in this frame $g_{\mu\nu}^{*}$ is defined by the conformal transformation $g_{\mu\nu}^{*} = \Phi g_{\mu\nu}$ and 
the action can be written in the form

\begin{eqnarray}
S = \frac{1}{16\pi G}\int d^4x\sqrt{-g^*}\left[R^* - 2g^{*\mu\nu}\partial_{\mu}\phi\partial_{\nu}\varphi  - V(\varphi)\right] + S_{Matter}(e^{-\frac{2}{\sqrt{3}}\phi}g_{\mu\nu}^*\chi),
\end{eqnarray}
where $R^*$ is the scalar curvature with respect  to the Einstein frame metric $g_{\mu\nu}^*$. 
The new scalar field is defined by $\varphi = \frac{\sqrt{3}}{2}\ln\Phi$, the potential in the Einstein frame is given by $V(\varphi) = A^4(\varphi)U(\Phi(\varphi))$ 
and $A^2(\varphi) = \Phi^{-1}(\varphi)$. The field equations in this frame are much simpler and that is why we will use it. Since the physical quantities are measured in the Jordan frame, we will make the transition between the two frames where necessary.

In our studies we will consider oscillations of static neutron stars. But in some of the asteroseismology relations the moment of inertia is used. A natural and straightforward way to define this quantity is via the slow rotation approximation. For this reason we will briefly examine here the more general framework of slowly rotating neutron stars in $R^2$ gravity.

The line element in a stationary and axisymmetric spacetime, keeping only first-order terms in the angular velocity $\Omega$, can be written in the form:

\begin{eqnarray}
ds^2_{*}= - e^{2\phi(r)}dt^2 + e^{2\Lambda(r)}dr^2 + r^2(d\theta^2 +
\sin^2\theta d\vartheta^2 ) - 2\omega(r,\theta)r^2 sin^2\theta  d\vartheta dt.
\end{eqnarray}

The explicit form of the field equations in the Einstein frame  is as follows

\begin{eqnarray}
&&\frac{1}{r^2}\frac{d}{dr}\left[r(1- e^{-2\Lambda})\right]= 8\pi G
A^4(\varphi) \rho + e^{-2\Lambda}\left(\frac{d\varphi}{dr}\right)^2
+ \frac{1}{2} V(\varphi), \label{eq:FieldEq1} \\
&&\frac{2}{r}e^{-2\Lambda} \frac{d\phi}{dr} - \frac{1}{r^2}(1-
e^{-2\Lambda})= 8\pi G A^4(\varphi) p +
e^{-2\Lambda}\left(\frac{d\varphi}{dr}\right)^2 - \frac{1}{2}
V(\varphi),\label{eq:FieldEq2}\\
&&\frac{d^2\varphi}{dr^2} + \left(\frac{d\phi}{dr} -
\frac{d\Lambda}{dr} + \frac{2}{r} \right)\frac{d\varphi}{dr}= 4\pi G
\alpha(\varphi)A^4(\varphi)(\rho-3p)e^{2\Lambda} + \frac{1}{4}
\frac{dV(\varphi)}{d\varphi} e^{2\Lambda}, \label{eq:FieldEq3}\\
&&\frac{dp}{dr}= - (\rho + p) \left(\frac{d\phi}{dr} +
\alpha(\varphi)\frac{d\varphi}{dr} \right), \label{eq:FieldEq4} \\
&&\frac{e^{\Phi-\Lambda}}{r^4} \frac{d}{dr}\left[e^{-(\Phi+ \Lambda)}r^4 \frac{d{\bar\omega}(r)}{dr} \right] =
16\pi G A^4(\varphi)(\rho + p){\bar\omega}(r) ,
\end{eqnarray}
where we have defined

\begin{eqnarray}
\alpha(\varphi)= \frac{d\ln A(\varphi)}{d\varphi} \;\;\; {\rm and}\;\;\;  \bar\omega = \Omega - \omega.
\end{eqnarray}
Here $\rho$ and $p$  are the energy and pressure density in the Jordan frame and they are connected to the Einstein frame ones via $\rho_{*}=A^{4}(\varphi)\rho$ and $p_{*}=A^{4}(\varphi) p$.

The above system of equations,  combined with the equation of state for the matter and appropriate boundary conditions, describes the interior  and the exterior of a compact star.
The boundary conditions are the natural ones. At the center of the star we have $\rho(0)=\rho_{c}, \Lambda(0)=0,\frac{d\varphi}{dr}(0) = 0, $ 
$\frac{d{\bar\omega}}{dr}(0)= 0$, where $\rho_{c}$ is a free parameter denoting the central value of the density.   
The condition for $\frac{d{\bar\omega}}{dr}(0)= 0$ ensures the regularity of the metric function $\bar\omega$ at the center of the star, 
the condition $\frac{d\varphi}{dr}(0) = 0$ ensures the regularity of the scalar field $\varphi$, and in turn the regularity of the Jordan frame scalar 
$\Phi$ at the center of the star. The regularity of the metric functions at $r=0$ requires $\Lambda(0)=0$ and since the Einstein and the Jordan 
frame metrics are conformally equivalent, this condition ensures also the regularity of the Jordan frame geometry at $r=0$. 
The boundary conditions at infinity are related to the fact that we consider asymptotically flat space-time. At infinity we have 
$\lim_{r\to \infty}\phi(r)=0,$  $\lim_{r\to \infty}{\bar\omega}=\Omega $ and $\lim_{r\to \infty}\varphi (r)=0$ with $V(0)=0$.
These conditions ensure the asymptotic flatness for both frames.

Although the above equations are in the Einstein frame, the final results we present in this work are in the Jordan frame (the physical one).
The coordinate radius $r_S$ of the star is determined by  $p(r_S)=0$ while  the physical radius in the Jordan frame is given by $R_{S}= A[\varphi(r_S)] r_S$.

The explicit  form of the conformal factor $A(\varphi)$ and the potential $V(\varphi)$ for $R^2$ gravity are
\begin{eqnarray}
A^2(\varphi)=
e^{-\frac{2}{\sqrt{3}}\varphi}, \;\;\;\alpha =- \frac{1}{\sqrt{3}},\;\;\; V(\varphi)= \frac{1}{4a}
\left(1-e^{-\frac{2\varphi}{\sqrt{3}}}\right)^ 2.
\end{eqnarray}

The moment of inertia $I$ of the  compact star is defined in the standard way

\begin{eqnarray}
I=\frac{J}{\Omega},
\end{eqnarray}
where J is the angular momentum of the start. More convenient for numerical calculations is to use an integral expression for the  moment of inertia,

\begin{eqnarray}\label{eq:I_integral}
I= \frac{8\pi G}{3} \int_{0}^{r_S}A^4(\varphi)(\rho + p)e^{\Lambda - \Phi} r^4 \left(\frac{\bar\omega}{\Omega}\right) dr .
\end{eqnarray}

From now on we shall use the dimensionless parameter $a\to a/R^2_{0}$  and the dimensionless moment of inertia
$I\to I/M_{\odot}R^2_{0} $ where $M_{\odot}$ is the solar mass and $R_{0}$ is one half the solar gravitational radius   $R_{0}=1.47664 \,{\rm km}$.

\section{Neutron and strange star oscillations in Cowling approximation}

In this section we will present the equations governing the non-radial oscillations of spherically symmetric compact stars in the Cowling approximation for $R^2$ gravity. In this case we are investigating  fluid perturbations on a fixed Jordan frame metric  which is equivalent to fixed scalar field and metric in the Einstein frame. Despite these simplifications the approximation turns out to be accurate enough for qualitative investigation. The results differ from GR in the range of 10 to 30\%  and the deviation decreases with the increase of the compactness \cite{Robe_68,Finn_88}.

The Jordan frame equations describing the perturbations in the Cowling formalism are obtained by varying the equation for the conservation of the energy-momentum tensor  
in the Jordan frame, namely $ \nabla_{\mu}\delta T^{\mu}_{\nu}=0$. The Jordan frame  Lagrangian fluid displacement  vector  $\vec\zeta$
can be parameterize by two  functions $W$ and $V$ in the standard  form \cite{Sotani_04}

\begin{eqnarray}
\vec\zeta = \left( e^{-{\tilde \Lambda}}W, - V\frac{\partial}{\partial \theta},
 - \frac{V}{\sin^2\theta}\frac{\partial}{\partial \vartheta} \right) \frac{e^{i\omega t}}{{\tilde r}^2}Y_{lm}(\theta,\vartheta),
\end{eqnarray}
where the Jordan frame metric function $e^{\tilde \Lambda}$ and the Jordan frame radial coordinate ${\tilde r}$ are given by
 $e^{\tilde \Lambda}=A(\varphi)e^{\Lambda}$ and ${\tilde r}=A(\varphi) r$. Equivalently $\vec\zeta$ can be written in the form

\begin{eqnarray}
\vec\zeta = A^{-2}(\varphi)\left(A(\varphi) e^{-\Lambda}W, - V\frac{\partial}{\partial \theta},
 - \frac{V}{\sin^2\theta}\frac{\partial}{\partial \vartheta} \right) \frac{e^{i\omega t}}{r^2}Y_{lm}(\theta,\vartheta).
\end{eqnarray}
Then the system of ordinary differential equations describing the star oscillations in Cowling approximation is the following:
\begin{eqnarray} \label{eq:WV_eq}
\frac{dW}{dr} &=& \frac{d\rho}{dp}\left(\omega^2A(\varphi)e^{\Lambda-2\Phi}Vr^2 + \Phi'W + \alpha(\varphi)\psi W \right) - l(l+1)A(\varphi)e^{\Lambda}V,\\
\frac{dV}{dr} &=& \left(2\left(\Phi' + \alpha(\varphi)\psi\right) - \frac{l}{r}\right)V - \frac{e^{\Lambda}WA^{-1}(\varphi)}{r^2} \nonumber,
\end{eqnarray}
where we have defined $\psi = d\varphi/dr$.

The boundary condition at the star surface is that the Lagrangian perturbation of the
pressure vanishes, which is equivalent to

\begin{eqnarray}
\left. \omega^2 e^{-2\Phi}V + \left(\Phi' + \alpha(\varphi)\psi\right)e^{-\Lambda}W\frac{A^{-1}(\varphi)}{r} = 0 \right|_{r=R}.
\end{eqnarray}
From a numerical point of view it is convenient to introduce new variables $W_1$ and $V_1$ and express the boundary condition at the center of the star in terms of them

\begin{eqnarray}
W=W_1r^{l+1}, \quad V=V_1r^{l}.
\end{eqnarray}

We are rewriting the system (\ref{eq:WV_eq}) and the boundary condition at the surface in therms of $W_1$ and $V_1$. The new system of equations has the following form:

\begin{eqnarray} \label{eq:WV1_eq}
\frac{dW_1}{dr} &=& \frac{d\rho}{dp}\left[\omega^2A(\varphi)e^{\Lambda-2\Phi}V_1r + \Phi'W_1 + \alpha(\varphi)\psi W_1 \right] - \frac{l(l+1)A(\varphi)e^{\Lambda}V_1}{r},\\
\frac{dV_1}{dr} &=& \left[2\left(\Phi' + \alpha(\varphi)\psi\right) - \frac{l}{r}\right]V_1 - \frac{e^{\Lambda}W_1A^{-1}(\varphi)}{r}, \nonumber
\end{eqnarray}
with the boundary condition at the surface of the star

\begin{equation}
\left. \omega^2e^{-2\Phi}rV_1 + \left(\Phi' + \alpha(\varphi)\psi\right)e^{-\Lambda}W_1A^{-1} = 0 \right|_{r=R}
\end{equation}
and at the center

\begin{equation}
\left. W_1 = -lA(\varphi)V_1 \right|_{r=0}.
\end{equation}

The above system (\ref{eq:WV1_eq}) combined with the boundary conditions forms an eigenvalue problem with the frequencies $\omega$ being the eigenvalues of the system.
In the next section we present the results from the numerical solution of the eigenvalue problem. 

\section{Results}

\begin{figure}[]
\centering
\includegraphics[width=0.48\textwidth]{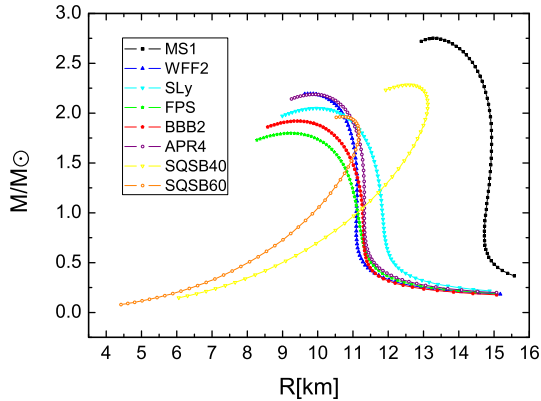}
\caption{The mass-of-radius relations for all EOS in pure GR. It could be seen that they cover a wide range of stiffnesses.}
\label{Fig:All_M(R)}
\end{figure}

We investigate numerically the fundamental mode (f-mode) frequencies obtained from the system of  eqs. (\ref{eq:WV1_eq}) supplied with six realistic hadronic equations of state and two quark ones. The eigenvalue problem is solved using a shooting method where the oscillation frequency plays the role of the shooting parameter. The two classes of equations of state cover a very wide range of masses and radii as shown in Fig. \ref{Fig:All_M(R)}. EOS FPS and BBB2 are the softest ones with masses below the observational limit of $2M\astrosun$ \cite{Demorest_10}.  Nevertheless we use them in order to check the EOS independence of the relations we derive \footnote{It should  be mentioned that in $f(R)$ theory of gravity some soft EOS could be reconciled with the observations because the maximum mass increases, as demonstrated in \cite{Yazad_14}.}. On the other hand EOS SLy, APR4 and WFF2 have masses above $2M\astrosun$ and fall into the preferred range of radii \cite{Lattimer2012}. EOS MS1 is the stiffest one with maximum mass over $2.7M\astrosun $ and much bigger radii.

For the hadronic EOS we use piecewise polytropic approximations \cite{Read_PPA}. The quark ones on the other hand have the analytical form

\begin{equation}
p = b(\rho - \rho_0),
\end{equation}
where the constants $b$ and $\rho_0$ are taken from \cite{Dorota08} for EOS SQS B60 and SQS B40. The first one gives masses slightly below two solar masses and the second one is stiffer and has higher masses and radii.

\begin{figure}[]
\centering
\includegraphics[width=0.48\textwidth]{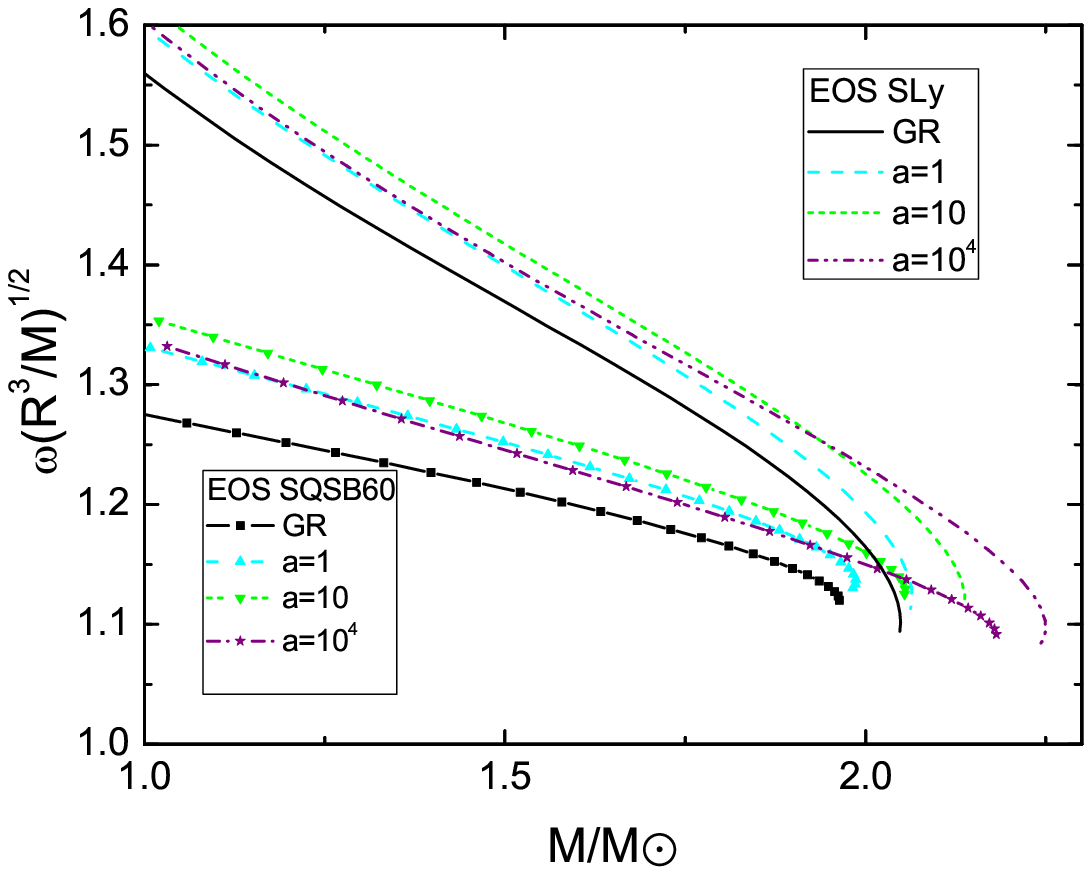}
\includegraphics[width=0.48\textwidth]{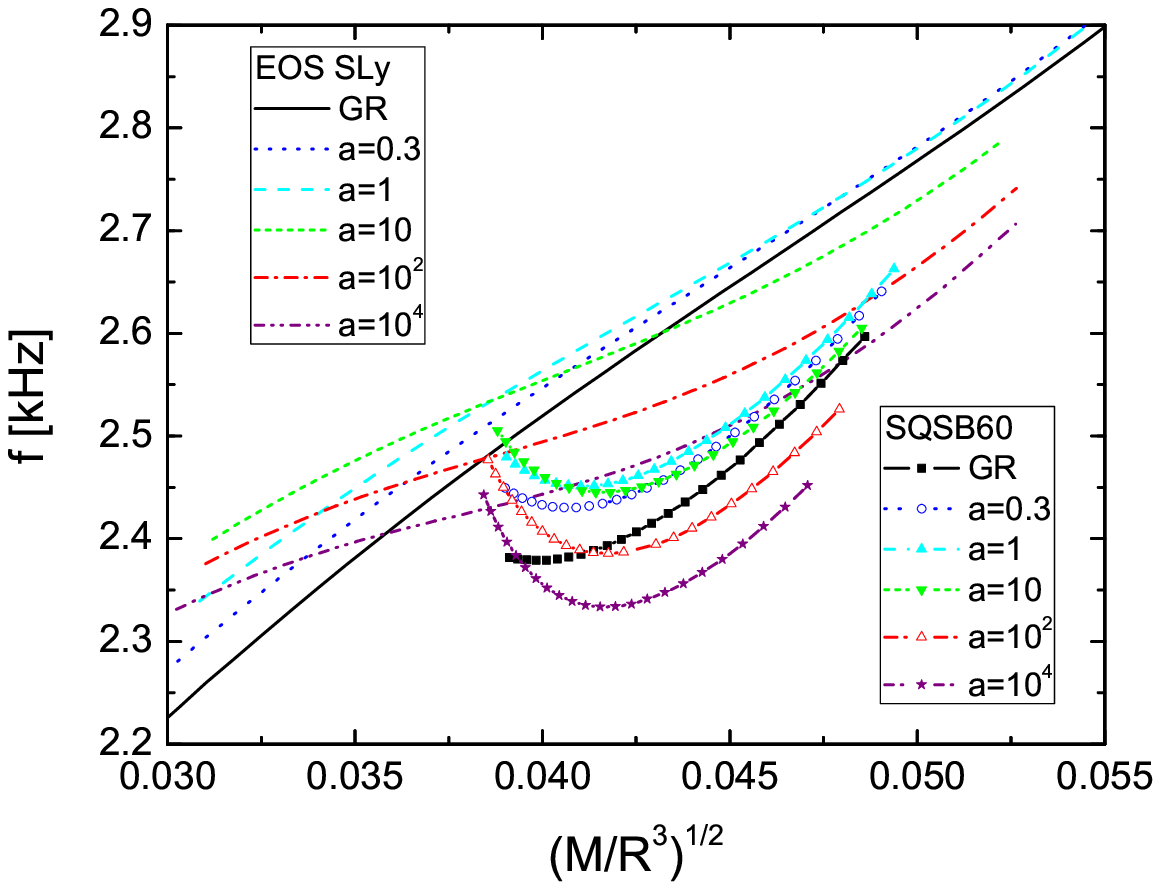}
\caption{Results for one hadronic EOS, namely SLy, and one quark EOS, namly SQSB60. Both have maximum mass around two solar masses. In the left panel the normalized frequency as a function of the neutron star mass is presented  while in the right panel -- the oscillation frequency in ${\rm kHz}$ as a function of the average density of the star in ${\rm km}^{-1}$ .}
\label{Fig:omega_freq}
\end{figure}

We will examine a wide range of relations between the $f$-mode oscillation frequencies and the stellar parameters such as mass, radius and moment of inertia. The particular choice is motivated by the most commonly used relations in the literature on gravitational wave asteroseismology. First we will investigate the neutron and strange stars oscillation frequencies and the differences between GR and $R^2$ gravity. We thoroughly investigated the f-modes for all EOS, but here we present results for one hadronic and one strange matter EOS, namely EOS SLy and SQS B60, that can be considered as representative examples. The sequences of models range from one solar mass up to the maximum one in accordance with the observations. In Fig. \ref{Fig:omega_freq} we present and compare the results for GR and $R^2$ gravity for neutron and strange stars. In the left panel the normalized frequency $\omega (R^3/M)^{1/2}$ as a function of the mass of the star, measured in solar masses, is presented for EOS SLy and SBSB60. Such a relation was used in other studies of alternative neutron star models \cite{Sotani_04,Sotani_09,Yazad_12,Doneva_12}. For clarity, the SQSB60 results are marked with additional symbols. The GR case is presented by a continuous black line and different values of $a$ are in different colors and patterns. It can be seen that for both GR and $f(R)$ gravity the results for neutron stars are much steeper than those for strange stars and the normalized frequency decreases with the increase of the mass.  The maximum  deviation from GR is below 10\% and the behavior is qualitatively different for the different mass ranges. For smaller masses the deviation increases when $a\leq 10$ but it starts to decrease for larger $a$. For bigger masses though the deviation increases monotonically with the increase of $a$.

In the right panel of Fig. \ref{Fig:omega_freq} we present the dimensional frequency $f = \omega/(2\pi)$, measured in kHz as a function of the average density of the star $(M/R^3)^{1/2}$ in ${\rm km}^{-1}$. This is one of the most standard asteroseismology relations used in the literature, originating from \cite{A_K_08}. A qualitative difference between the results for neutron stars in GR and in $f(R)$ theory can be seen on the graph. The dependences in $f(R)$ theories are no longer linear for large values of $a$. For strange stars the graphs have similar shapes for both theories.
For small values of $a$ the mode frequencies in $R^2$ gravity are always higher than GR, and a significant decrease with respect to  GR is observed for larger average densities and large  values of the parameter $a$.

\begin{figure}[]
\centering
\includegraphics[width=0.48\textwidth]{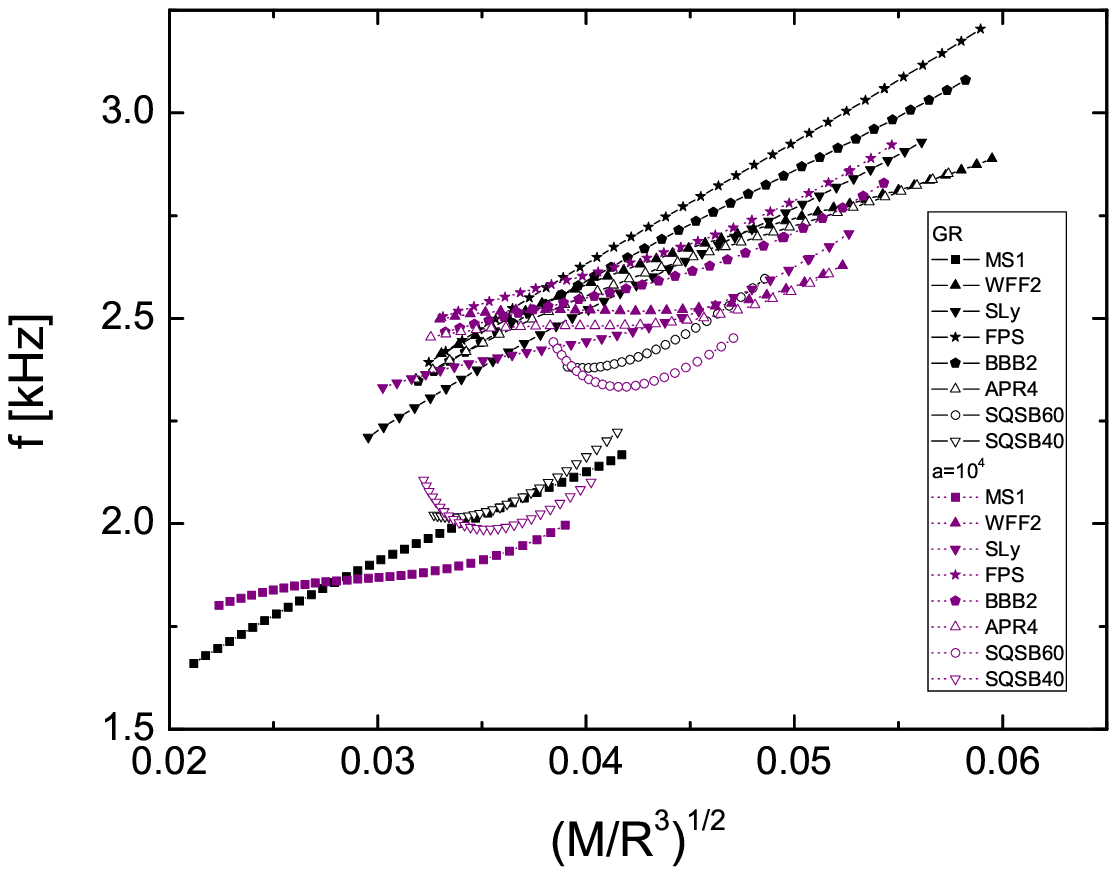}
\includegraphics[width=0.48\textwidth]{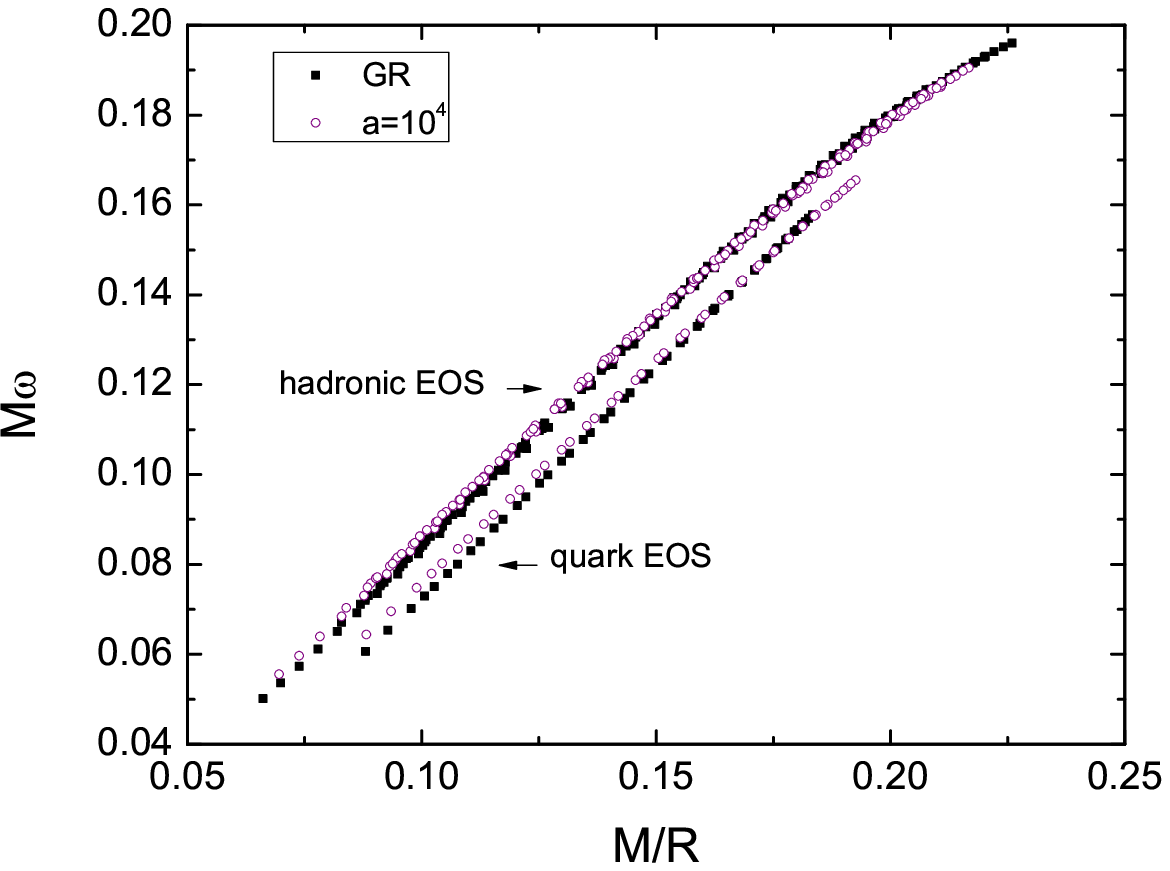}
\caption{In the left panel we present the frequency as a function of the average density for all examined EOS for the case of GR and for $ R^2$ gravity with parameter $a=10^4$, which induces a deviation close to the maximum possible. It could be seen that the results for the stiffest EOS, namely MS1 and SQSB40, seriously differ from all the other results. However, the two groups of results form two narrow bands. In the right panel we plot the relation $M\omega(M/R)$. The presented results are again for all EOS for GR and $f(R)$ theory with $a=10^4$. As one can see, the results for  hadronic and quark EOS are only shifted from each other, but there is quite good universality for both cases independently. The separation between the GR and $f(R)$ dependencies is  negligible.    }
\label{Fig:f_omegaM/R}
\end{figure}

In the left panel of Fig. \ref{Fig:f_omegaM/R}   a graph similar to the one in the right panel of Fig. \ref{Fig:omega_freq} is presented but in this one all EOS are included for GR and $R^2$ gravity with $a=10^4$. This value of $a$ is chosen because our investigations show that it gives almost the maximum possible deviation from the pure Einstein's theory. Different EOS are marked with different symbols, the GR case is in black and $a=10^4$ is in color. Two groups can be distinguished in the graph -- one for the softer and the typical EOS (with maximum mass around and below $2M\astrosun$), and one for stiffer EOS  (contains EOS MS1 and SQS B40). We will concentrate on the former group which could be separated into two bands, one formed by the results for GR and the other for $a=10^4$. These two bands fully overlap for small values of the average density, and partially for higher densities. We should note that strange stars do not fit very well in the band formed by neutron star solutions in Fig. \ref{Fig:f_omegaM/R}  \cite{Benhar_04,Lau_10}. Let us comment on the deviations coming from the modification of GR. Although the difference between GR and $R^2$ gravity is non-negligible, i.e. up to 10\%, it is comparable with the uncertainty of the EOS. Constraining further the EOS by different observations will reduce the spread of the data due to the EOS uncertainty. This will make the difference between the two theories of gravity clearer and that can be potentially used to set constraints on the $f(R)$  theories using future gravitational wave observations.

The empirical relations describing these dependences have been widely investigated in the literature starting from  \cite{A_K_08}. They have used linear fitting for the frequency-average density relation and we will employ the same. We have excluded from the relation the very stiff nuclear EOS MS1 and the quark ones SQSB60 and SQSB40 since they clearly lead to significantly different dependences and they are not favored by  observations. The results for  GR and for $R^2$ gravity with $a=10^4$ are fitted separately using a linear fit of the form
\begin{equation} \label{eq:lin_fit}
f = C_1 + C_2\sqrt{\frac{M}{R^3}}.
\end{equation}
The dimensions of the constants for this fit, $C_1$ and $C_2$, are as follows: $C_1$ is in ${\rm kHz}$ and $C_2 $ is in ${\rm kHz / km}$. For the case of GR we have $C_1 = 1.59, C_2 = 24.23$, and for $R^2$ gravity with $a=10^4$ we have $C_1 = 1.95, C_2 = 14.25$. From these numbers it is obvious that the results in GR are much steeper than these in $f(R)$.

Let us proceed to the next asteroseismology relations we plan to consider, where different normalization of the quantities is used. In the right panel of Fig.  \ref{Fig:f_omegaM/R} we present the scaled frequency $M\omega$  as a function of the compactness of the star $M/R$, as proposed in
\cite{A_K_08}. The graph contains all EOS, the GR solutions are marked with black squares and the ones for $a=10^4$ with circles.
As one can see normalizing the frequency seriously decreases the spread of the data due to the EOS and therefore it leads to significant EOS independence \cite{A_K_08, Benhar_04,Tsui_04}.  The difference between GR and $R^2$ gravity is below 5\% which is not large enough to be observable.

For small values of $M/R$ the relation is almost linear, but close to the maximum values of the compactness this changes. That is why we use a cubical polynomial fit in this case,

\begin{equation} \label{eq:cub_fit}
M\omega = C_1 + C_2\frac{M}{R} +  C_3\left(\frac{M}{R}\right)^2  +  C_4\left(\frac{M}{R}\right)^3.
\end{equation}
We excluded both quark EOS and fitted the rest of our data for three different cases. For the pure GR case $C_1 = 4.95\times10^{-3}, C_2 = 4.16\times10^{-1}, C_3 = 5.17, C_4 = -14.43$, and for the $R^2$ gravity $C_1 = -2.54\times10^{-3}, C_2 = 6.48\times10^{-1}, C_3 = 3.67, C_4 = -11.57$. If one uses the data for both theories we have  $C_1 = 2.76\times10^{-4}, C_2 = 0.544, C_3 = 4.37, C_4 = -12.95$.

\begin{figure}[]
\centering
\includegraphics[width=0.48\textwidth]{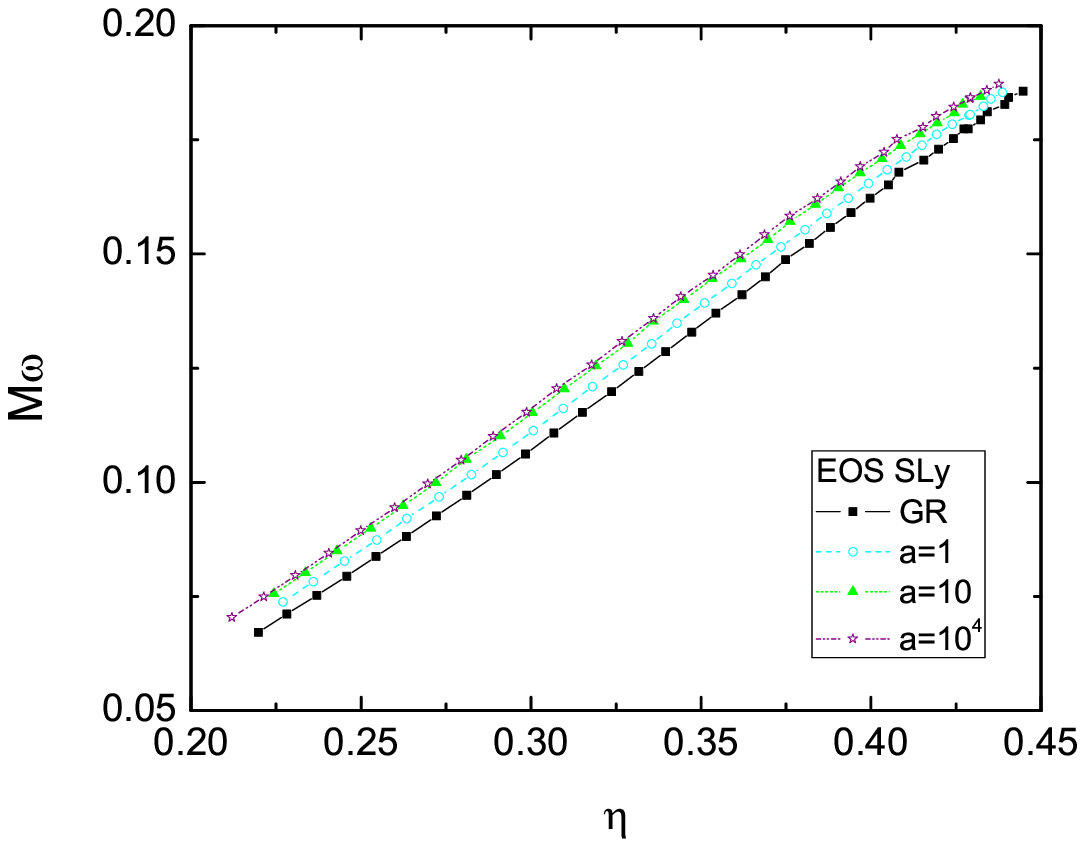}
\includegraphics[width=0.48\textwidth]{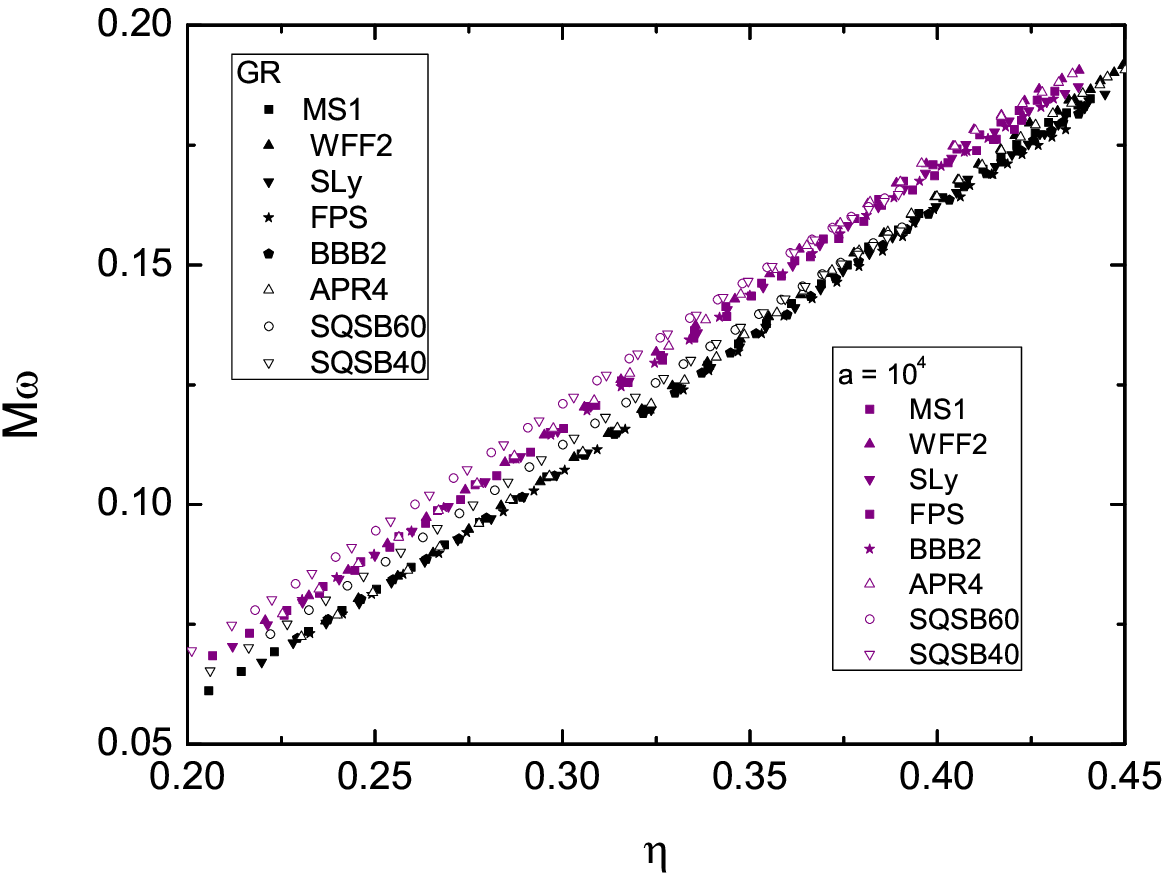}
\caption{The relation $M\omega(\eta)$ is presented in the graphs. In the right panel we investigate the deviations in this relation for EOS SLy and several different values of the parameter $a$. In the left panel results for all EOS are presented.The relation for both the neutron and the strange stars are quite EOS independent and a visible difference between the GR and the $a=10^4$ cases exists. The maximum observed deviation is close to 10\%. }
\label{Fig:uni}
\end{figure}

As one can see the results for strange stars are also quite EOS independent but they differ from the neutron star ones. More precisely they are shifted with some constant value of the normalized frequency. Lau et al. \cite{Lau_10} suggested that this is due to the difference in the density profiles of strange and neutron stars. That is why they used a new parameter for constructing the universal asteroseismology relations. Instead of the compactness they employ the parameter $\eta \equiv \sqrt{M^3/I}$, where $I$ is the moment of inertia of the star.
Therefore we have also investigated the relation $M\omega(\eta)$.
In the left panel of Fig \ref{Fig:uni} we present results for EOS SLy for GR and a few different value of the parameter $a$. The biggest deviation is around 10\% and it is relatively the same for all values of $a\geq 10$. In the right panel we present our results for all EOS  for GR and for $a=10^4$. As we pointed out in Fig. \ref{Fig:f_omegaM/R},  if we use the compactness $M/R$ as a dimensionless parameter and normalize $\omega$, the deviation between the two theories is under 5\%. Using $\eta$ as a parameter preserves the EOS independence but  the results for GR and $R^2$ gravity get further apart. However the deviation between GR and $R^2$ gravity  is still small, up to roughly 10\% as we commented. In this case the dependences for both theories are more or less shifted with some constant value of $M\omega$. As far as the difference between neutron and strange stars is concerned, it decreases a lot if we use $\eta$ as a dimensionless parameter instead of $M/R$, but the results in the two cases do not fully coincide as opposed to the ones in \cite{Lau_10}. This is most probably due to the fact that we are not solving the full nonlinear Einstein equations, but the Cowling approximation is employed instead.

In order to obtain asteroseismology dependences for these normalizations, we are using a quadratic fit similar to \cite{Lau_10},

\begin{equation} \label{eq:quad_fit}
M\omega = C_1 + C_2\eta +  C_3\eta^2.
\end{equation}
For GR we obtain $C_1 = -3.28\times10^{-2}, C_2 = 0.426, C_3 = 0.157$ and for $R^2$ gravity with $a=10^4$ we have $C_1 = -4.29\times10^{-2}, C_2 = 0.544, C_3 = -2.57 \times 10^{-2}$.

\section{Conclusions}

In this paper we investigated neutron and strange stars oscillations in GR and $R^2$ gravity. More precisely we concentrated on the f-mode oscillations since they are expected to be very efficient gravitational wave emitters. We employed a big range of hadronic and strange matter EOS with different stiffness and the calculations were performed in the Cowling approximation. The observed maximum deviation between the f-mode frequencies in  GR and $R^2$ gravity is up to 10\% and depends on the value of the $R^2$ gravity parameter $a$.

We investigated multiple gravitational wave asteroseismology relations available in the literature and obtained the corresponding analytical fits.
In most cases the dependences in GR and $R^2$ gravity are qualitatively the same -- they are only shifted with respect to one another. 
The only exception is the relation connecting the f-mode oscillation frequencies to the average density of the star where some qualitative 
differences between the two theories exist. As mentioned above these differences do not exceed 10\%. 
Such deviations are big enough to make a clear distinguishment between GR and $R^2$ gravity, but as far as the real gravitational wave observations are concerned, 
they will most probably be below the accuracy of the observed gravitational wave frequencies.

However this result also can be considered as a strong point of the asteroseismology relations examined in the present paper. The reason is that one can make the conclusion that the relations we consider turned out to be not only EOS independent, but up to a large extent theory independent too. Therefore solving the inverse problem will supply us with unique values for the parameters of a star, like mass and radius, insensitive to both the EOS and the particular theory of gravity. In order to prove this conjecture more rigorously one of course has to examine these relations is other alternative theories of gravity, but the present results give a hint in this direction.

\section*{Acknowledgements}

DD would like to thank the Alexander von Humboldt Foundation for a stipend. KK, KS and  SY would like to thank the
Research Group Linkage Programme of the Alexander von Humboldt Foundation for the support. The support by
the Bulgarian NSF Grant DFNI T02/6 and "New-CompStar" COST Action MP1304 is gratefully acknowledged.



\begin{thebibliography}{99}

\bibitem{Sotiriou_10} T. P. Sotiriou and V. Faraoni, Rev.Mod.Phys. 82, 451 (2010)

\bibitem{De Felice_10} A. De Felice and S. Tsujikawa, Living Rev.Rel. 13, 3 (2010)

\bibitem{Nojiri_11} S. Nojiri and S. D. Odintsov, Phys.Rept. 505, 59 (2011)


\bibitem{A_K_06} N. Andersson, K. Kokkotas, Phys. Rev. Lett. 77, 4134 (1996)

\bibitem{A_K_08} N. Anderson, K. Kokkotas,  Mon. Not. Roy. Astron. Soc. 299, 1059 (1998)

\bibitem{Benhar_04} O. Benhar, V. Ferrari, L. Gualtieri, Phys. Rev. D70, 124015 (2004)

\bibitem{Lau_10} H.K. Lau, P.T. Leung, L.M. Lin,  ApJ 714, 1234 (2010)

\bibitem{Chirenti_15} C. Chirenti, G. H. de Souza, W. Kastaun, arXiv:1501.02970 [gr-qc]

\bibitem{Sotani_03} H. Sotani, T. Harada, Phys. Rev. D 68, 024019 (2003)

\bibitem{Sotani_09} H. Sotani, Phys. Rev. D80, 064035 (2009)

\bibitem{Yazad_12} S. Yazadjiev and D. Doneva JCAP 03, 037 (2012)

\bibitem{Doneva_12} D. Doneva, S. Yazadjiev, Phys. Rev. D 85, 124023 (2012)

\bibitem{Sotani_14} H. Sotani, Phys. Rev. D89, 124037 (2014)

\bibitem{Krueger_14} C. J. Krüger, W. C. G. Ho, N. Andersson,  arXiv:1402.5656 [gr-qc]

\bibitem{Sotani_04} H. Sotani, K. Kokkotas,  Phys. Rev. D 70, 084026 (2004)

\bibitem{Sotani_2005} H. Sotani and K. Kokkotas, Phys. Rev. D 71, 124038 (2005)

\bibitem{Sotani_10} H. Sotani, Phys. Rev. D82, 124061 (2010)

\bibitem{Silva_2014}  H.~Silva, H.~Sotani, E.~Berti and M.~Horbatsch, Phys.\ Rev.\ D {\bf 90}, 124044 (2014)

\bibitem{Gaertig_11} E. Gaertig, K. Kokkotas, Phys. Rev. D 83, 064031 (2011)

\bibitem{Doneva_13} D. Doneva, E. Gaertig, K. Kokkotas, C. Krüger, Phys. Rev. D 88, 044052 (2013)

\bibitem{Kokkotas_2001} K. Kokkotas, T. Apostolatos, N. Andersson, MNRAS 320, 307(2001)

\bibitem{Lattimer_05} M. Lattimer, B. Schutz, ApJ, 629, 979 (2005)

\bibitem{Yagi_13} K. Yagi and N. Yunes, Science 341, 365 (2013).
\bibitem{Yagi_13a} K. Yagi and N. Yunes, Phys. Rev. D 88, 023009 (2013).
\bibitem{Maselli_13} A. Maselli, V. Cardoso, V. Ferrari, L. Gualtieri, and P. Pani, Phys. Rev. D 88, 023007 (2013).
\bibitem{Doneva_14a} D. Doneva, S. Yazadjiev, N. Stergioulas, and K. Kokkotas, ApJ 781, L6 (2014)
\bibitem{Pappas_14} G. Pappas and T. A. Apostolatos, Physical Review Letters 112, 121101 (2014)
\bibitem{Chakrabarti_14} S. Chakrabarti, T. Delsate, N. G¨ urlebeck, and J. Steinhoff, Physical Review Letters 112, 201102 (2014)
\bibitem{Haskell_14} B. Haskell, R. Ciolfi, F. Pannarale, L. Rezzolla, MNRAS Letters 438, L71-L75 (2014)
\bibitem{Urbanec_13} M. Urbanec, J. C. Miller, and Z. Stuchlik, Mon. Not. Roy. Astron. Soc. 433, 1903 (2013)

\bibitem{Yazad_14}  S. Yazadjiev, D. Doneva, K. Kokkotas, K. Staykov,  JCAP  06, 003 (2014)

\bibitem{Staykov_14}   K. Staykov, D. Doneva, S. Yazadjiev, K. Kokkotas, JCAP, 1410, 006 (2014)

\bibitem{Robe_68} H. Robe, Annales d?Astrophysique 31, 475 (1968).
\bibitem{Finn_88} L. S. Finn, Mon. Not. Roy. Astron. Soc. 232, 259 (1988).

\bibitem{Yazad_15} S. Yazadjiev, D. Doneva, K. Kokkotas, arXiv:1501.04591 [gr-qc]

\bibitem{Demorest_10} P. Demorest, T. Pennucci, S. Ransom, M. Roberts and J. Hessels, Nature 467, 1081 (2010)

\bibitem{Lattimer2012} J. M. Lattimer, Annu. Rev. Nucl. Part. Sci. 62 (2012) 485.
\bibitem{Read_PPA} J. Read, B. Lackey, B. Owen, and J. Friedman Phys. Rev. D 79, 124032 (2009)

\bibitem{Dorota08} D. Gondek-Rosinska and F. Limousin, Arxiv e-prints (2008), 0801.4829.

 \bibitem{Tsui_04} L. K. Tsui, P. T. Leung, Mon. Not. Roy. Astron. Soc. 357,1029 (2005)


\end{thebibliography}

\end{document}